Marta Magdalena Szymczyk *

# ANALYSIS OF FLEXIBLE TRAFFIC CONTROL METHOD IN SDN

*Abstract:* The aim of this paper is to analyze methods of flexible control in SDN networks and to propose a self-developed solution that will enable intelligent adaptation of SDN controller performance. This work aims not only to review existing solutions, but also to develop an approach that will increase the efficiency and adaptability of network management. The project uses a modern type of machine learning, Reinforcement Learning, which allows autonomous decisions of a network that learns based on its choices in a dynamically changing environment, which is most similar to the way humans learn. The solution aims not only to improve the network's performance, but also its flexibility and real-time adaptability - flexible traffic control.

## 1. Introduction

One of the most advanced solutions that have emerged in response to the challenges of modern computer networks are Software-Defined Networks (SDN) [1]. They enable efficient management of resources and network traffic, a definite advantage in the age of increasingly complex networks requiring dynamic management. By centralizing control and enabling flexible management, SDN offers new opportunities for network optimization. Nevertheless, fully realizing the potential of SDN requires the development of advanced and adaptive control methods.

This article focuses on analyzing current methods of flexible control for SDN networks and presenting a solution to improve the efficiency and adaptability of network management. The approach presented is based on the application of machine learning, specifically the Reinforcement Learning (RL) [2]. This technique allows networks to make autonomous decisions based on previous experiences and dynamically changing conditions, which is similar to the way humans learn.

The goal of the proposed solution is to not only increase network performance, but to improve its flexibility and real-time adaptability. The use of reinforcement learning enables dynamic and flexible control of network traffic, resulting in more efficient and responsive resource management [3]. The article reviews existing solutions and describes in detail the original approach developed in-its own, pointing out its potential benefits and implementation possibilities.

## 2. Project implementing neural network in SDN

### 2.1. General scheme and topology

The design part of the article presents the possibility of learning a neural network based on reinforcement learning, where the neural network searches for an optimal configuration that meets the requirements set for the SDN. The problem of finding the shortest path in the network is presented as an example of such an operation. Using the REST API from Floodlight [4], the structure and parameters of the network are loaded. Based on this data, a model for learning the neural network was created, then the network was subjected to learning, as a result of which it finds the optimal path in the network set for it. The final result of the neural network is then transformed into a configuration, which is sent back via REST API to Floodlight.

---

* AGH University of Krakow, ORCID (Marta Szymczyk) : https://orcid.org/0009-0007-0679-2407

The proposed system (Fig. 1) includes a neural network. Its task is to determine the shortest path by which traffic will be sent. It communicates with the SDN controller via the north interface, specifically the REST API. Communication between the SDN controller and the switches is possible via the south interface. The standard OpenFlow communication protocol was chosen for this. During the initial configuration of the project, a predefined network topology is sent to the SDN controller using the Mininet tool [5].

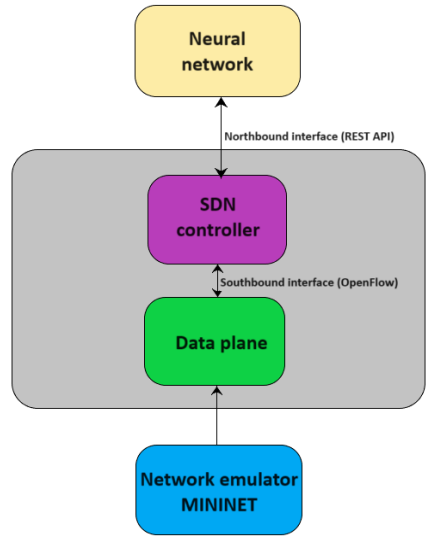

**Figure 1 Architecture of the system**

### 2.2. Implementation

#### 2.2.1. Network emulation - network configuration in Mininet

The network topology under test was run using the Mininet tool and a special command in which we specify as parameters that the SDN controller is not running locally, its IP address, the port on which the controller will communicate, the path under which the custom topology can be found, the use of the custom topology "mytopo", the use of the "Traffic Control" tool, the bandwidth of the links and the delay on the link.

#### 2.2.2. Data loading and preparation

Using a custom load-and-prepare script and a REST API, data from the SDN network is downloaded into the neural network environment. The script performs a series of operations to retrieve and process the network data managed by the SDN controller. The code efficiently processes the SDN network topology data, creates a matrix representation of the connections and identifies key end nodes in the network thus preparing the data for neural network operation.

### 2.2.3. Project and implementation of DQN

In the project, the environment for machine learning using the GYM [6] library is defined. At the beginning, the correctness of the input data received from SDN and needed for learning the neural network is checked and the whole environment is prepared. During the learning process, a record of this process will be created in the form of memorizing the probabilities of the model's actions. For this purpose, a one-dimensional vector is created for the starting node, and then the predicted probabilities for each action are saved to a CSV file after each learning epoch.

An important part of the implementation is the "create_model" function defined in the project, shown in Fig. 2, which creates a neural network model. Three layers are defined in the model, including two hidden layers with ReLU activation, and a third output layer with softmax activation, which converts the output to probabilities summing to 1. It uses the Adam optimizer and the loss function "categorical_crossentropy," a standard loss function for multiclass classification problems. The function's arguments are the number of input features, or input dimensions, the number of output classes, or output dimensions, and the learning factor for the Adam optimizer, set by default to 0.01. A sequential model is created to add layers in a linear fashion. The function returns the configured model.

```python
def create_model(input_dim, output_dim, learning_rate=0.01):
    model = keras.Sequential(
        [
            layers.Dense(64, input_dim=input_dim, activation='relu'),
            layers.Dense(64, activation='relu'),
            layers.Dense(output_dim, activation='softmax')
        ])
    model.compile(optimizer=Adam(learning_rate=learning_rate),
                  loss='categorical_crossentropy')
    return model
```

**Figure 2 Creating the model of neural network**

A very important step in the project is to train the neural network model in the environment (Fig. 3). The argument of the function is the environment and the number of training episodes, set to 1000 by default. At the beginning, variables corresponding to the number of nodes in the network and the input and output dimensions, which are just equal to the number of nodes, are initialized. Next, a neural network model is created using the "learning_model" function. Using a loop, the function goes through each training episode, at the beginning of the loop the environment is reset to the initial state. The next step is to loop through the states in one episode. The loop continues until the current node is not the end node. The model predicts the probabilities for each action of the current state and stores them. Then the probabilities for unconnected nodes are reset to zero, and the modified probabilities are summed. If this sum is different from zero then the probabilities are normalized. The next node is selected based on the probabilities of the action. The reward is then calculated and the state is updated. The model is then trained for one epoch, after which the current state and node are updated. Finally, the function returns the trained model.

```python
def learning_model(env, num_episodes=1000):
    num_nodes = env.num_nodes
    input_dim = num_nodes
    output_dim = num_nodes
    model = create_model(input_dim, output_dim)

    for ep in range(num_episodes):
        print('.............. Episode: ', ep, ' ...............')
        state = env.reset()

        while env.current_node != env.end_node:
            action_probs = model.predict(state, verbose=0, workers=4,
                use_multiprocessing=True)[0]
            save_action_probs(model, env, ep)

            # It cannot stay at the same node or if there is no
            # connection between nodes
            tmp_action_probs = action_probs.copy()
            for i in range(num_nodes):
                if env.net[env.current_node, i] == 0:
                    tmp_action_probs[i] = 0
            sum_row = np.sum(tmp_action_probs)
            if sum_row != 0:
                tmp_action_probs /= sum_row
            else:
                print('Continue')
                env.current_node = env.end_node
                continue
            action_probs = tmp_action_probs.copy()

            # Action
            next_node = np.random.choice(range(num_nodes),
                                         p=action_probs)

            # Calculation of the reward
            reward = env.net[env.current_node, next_node]

            # Update of the state
            next_state = np.zeros((1, num_nodes))
            next_state[0, next_node] = 1

            # Training a model
            target = np.zeros((1, num_nodes))
            target[0, next_node] = reward

            print('FIT: current node: ', env.current_node,
                  ' next node: ', next_node)

            model.fit(state, target, epochs=1, verbose=0, workers=4,
                use_multiprocessing=True)

            state = next_state
            env.current_node = next_node

    return model
```

**Figure 3 Training neural network model**

Once this is done, a path from the start node to the end node is found in the simulation environment using a trained neural network model. Such a result is in turn prepared to be sent back to the SDN network.

### 2.2.4. Transfer of results to SDN

To send back the results of the neural network, an output script is used, which adds static flows to the SDN network based on the shortest path between nodes. The data obtained from the neural network is translated into the form needed to program (modify the configuration) of the computer network and sent to the SDN controller.

## 2.3. Test topology

The test case uses the topology shown in Figure 8, which is individually created using a script written in Python. The topology was chosen to represent how the proposed algorithm would handle a wide-area network problem. An important element in the topology is two main branches with cross-connection. Two hosts were used in the topology, as well as fourteen switches. Two of them were used for host connections, and the other twelve are connected to their nearest neighbors.

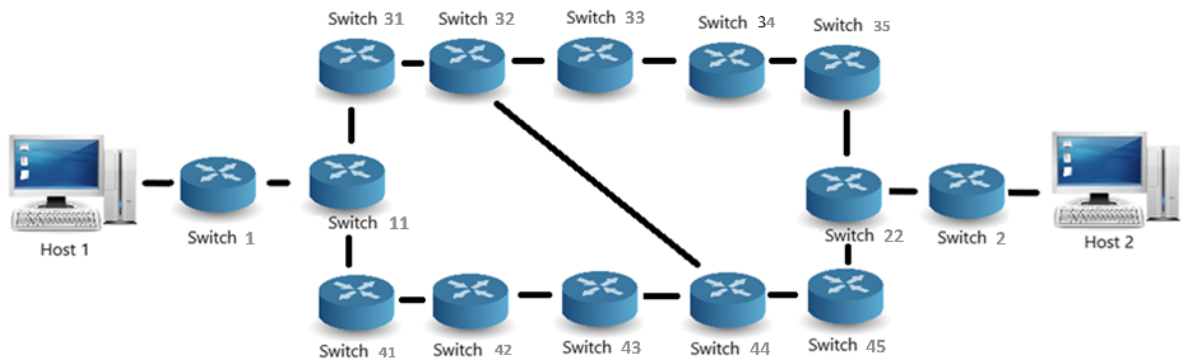

**Figure 4 Network topology**

Topology of used SDN:

- Number of hosts: 2
- Number of switches: 14
- Number of network connections: 17

Neural network parameters:

- Number of network layers: 3
- Network layers:
    - Dense with dimension 64 and activation function "relu"
    - Dense with dimension 64 and activation function "relu"
    - Dense with dimension 16 and activation function „softmax"
    - Parameter „learnin rate" = 0,01

o Loss function = „categorical_crossentropy"

## 2.4. Test results

The graphs shown in Fig. 5 and Tab. 1 show that the number of epochs after which the network has learned is 60.

**Table 1 "Action probes" after each epoch**

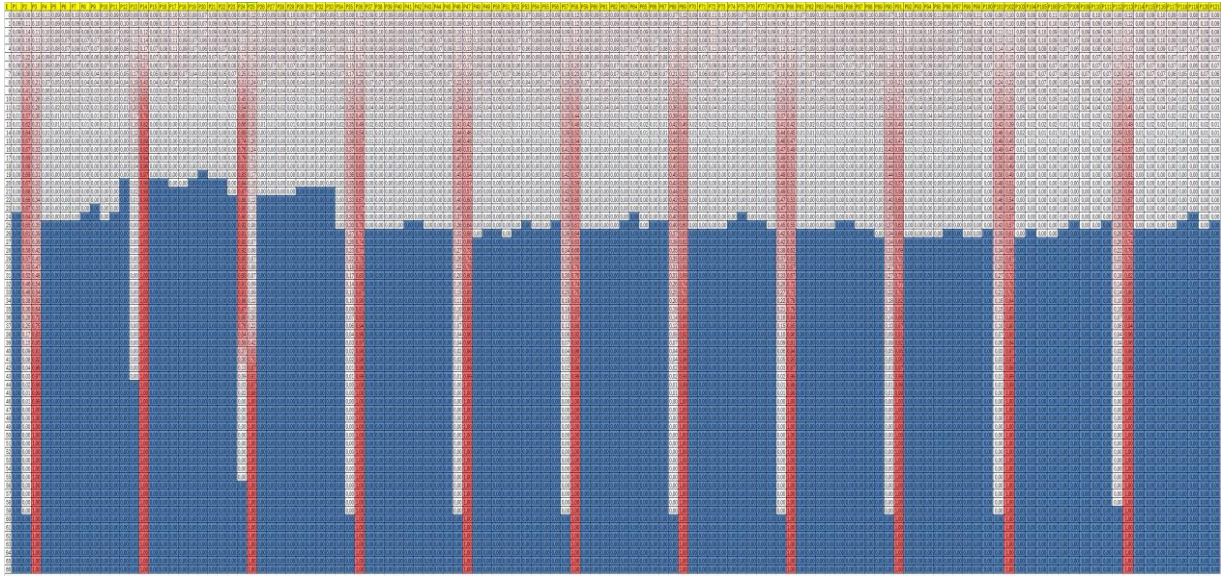

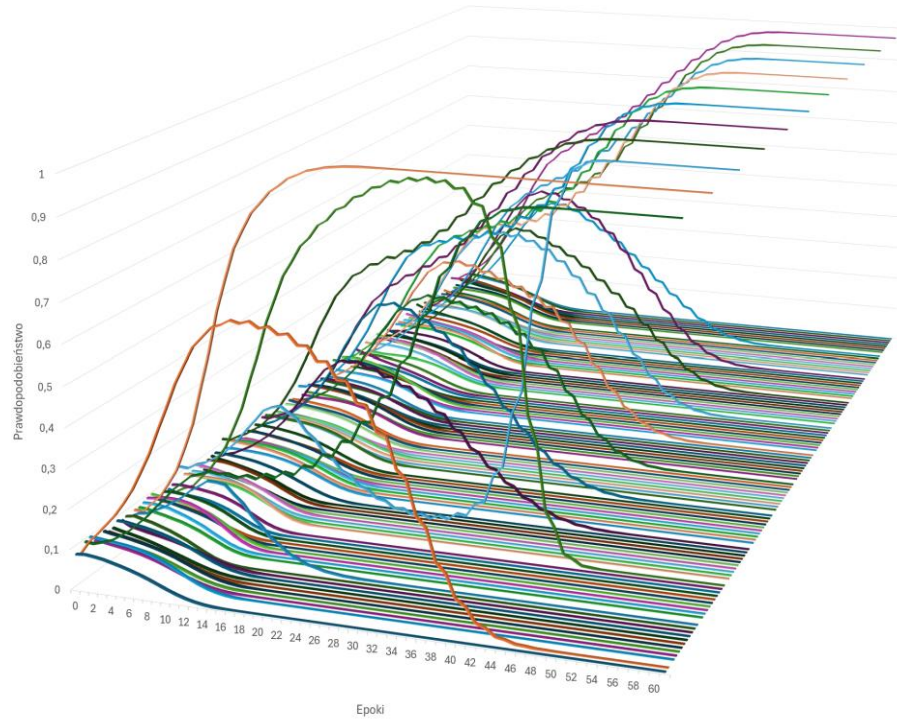

**Figure 5 A spatial graph of the "action probes" values after each epoch**
M. Szymczyk *Analysis of flexible traffic control method in SDN*
6

### 2.5. Interpretation of test results

In the case analyzed, the neural network exhibited stable behavior and had a fast learning process, which can be observed from the fact that it reached a saturation state after several tens of epochs. Based on this, it can be concluded that the selected parameters of the neural network were optimal and do not require adjustments. However, for larger sizes of SDN networks, some modifications to the parameters may be needed. Analysis of the values of the "action probes" shows that the initial values varied, and as the epochs progressed, the neural network pursued uniform results close to the value of 1, indicating effective learning and minimization of prediction errors. The results suggest that neural networks can be an effective tool for predicting and managing traffic in SDN networks. The model can be used for network monitoring, routing optimization and congestion prevention. Because of the model's rapid learning, neural networks can be applied in dynamic environments where network conditions can change rapidly. The model can adapt to new traffic patterns, which is crucial for efficient network management. Although the research was conducted on a relatively small network, the methods and results suggest that the technique can be scaled to larger and more complex SDNs, paving the way for further research and implementation in real production environments.

### 3. Conclusion

The aim of the study was to develop an in-house solution to intelligently adjust the operation of the SDN controller, which was achieved. The project used reinforcement learning, which allows autonomous decision-making by a network that learns from its choices in a dynamically changing environment, enabling flexible traffic control. The work aimed to improve the network's performance, flexibility and adaptability in real time.

In the case analyzed, the neural network exhibited stable behavior and had a fast learning process, which can be observed from the fact that it reached a saturation state after several tens of epochs. Based on this, it can be concluded that the selected parameters of the neural network were optimal and do not require adjustments. However, for larger SDNs, some modifications to the parameters may be needed.

The application of machine learning in SDN enabled new possibilities for flexible traffic control. The neural network efficiently and effectively found the optimal solution to the problems of given topologies. Machine learning is also applied in SDN to solve new problems not previously considered in SDN networks [7], as long as the user is able to define the problem and write it down in the form of guidelines for learning the network. The neural network is able to efficiently make SDN modifications by analyzing and finding the best solutions with changing requirements and dynamically changing state of SDN networks [8].

Future work can focus on further adjusting the parameters of neural networks for larger and more complex SDNs to further improve their performance and adaptability [9]. Thus, there is the possibility of making the networks even more flexible. The integration of other machine learning techniques into the proposed solution can also be explored to improve its efficiency. In addition, it is worth conducting tests in real network environments to verify the practical usability and scalability of the developed system.